\documentclass[12pt]{article}
\pdfoutput=1 
\usepackage{amsfonts}
\usepackage{amsmath,amssymb,amsthm}
\usepackage{graphicx}

\newcommand{\bea}{\begin{eqnarray}}
\newcommand{\eea}{\end{eqnarray}}

\newcommand{\half}{\frac{1}{2}}

\newcommand{\p}{\partial}

\newcommand{\scN}{\ensuremath{\mathcal{N}}}
\newcommand{\scO}{\ensuremath{\mathcal{O}}}

\def\SU{SU}
\def\U{U}

\newcommand{\be}{\begin{equation}}
\newcommand{\ee}{\end{equation}}
\newcommand{\beq}{\begin{eqnarray}}
\newcommand{\eeq}{\end{eqnarray}}
\newcommand{\unhat}{}

\begin{document}

\begin{titlepage}
\begin{flushright}
\normalsize
LPTENS-12/23 \\
PUPT-2418
\end{flushright}

\vfil

\bigskip

\begin{center}
\LARGE
Large $N$ Free Energy of 3d $\mathcal{N}=4$ SCFTs and AdS$_{4}$/CFT$_{3}$
\end{center}

\vfil
\medskip

\begin{center}
\def\thefootnote{\fnsymbol{footnote}}

Benjamin Assel$^{\natural}$, John Estes$^\flat$ and Masahito Yamazaki$^\sharp$

\end{center}

\begin{flushleft}\small

\vskip 5mm

\centerline{$^\natural$ Laboratoire de Physique Th\'eorique de l'\'Ecole Normale Sup\'erieure, }
\centerline{24 rue Lhomond, 75231 Paris cedex, France}

\vskip 5mm

\centerline{$^\flat$ Instituut voor Theoretische Fysica, KU Leuven, }
\centerline{Celestijnenlaan 200D B-3001 Leuven, Belgium}

\vskip 5mm

\centerline{$^\sharp$
Princeton Center for Theoretical Science, Princeton University, }
\centerline{Princeton, NJ 08544, USA}
\end{flushleft}

\bigskip

\thispagestyle{empty}

\setcounter{tocdepth}{2}

\bigskip

\begin{center}
{\bfseries Abstract}
\end{center}

We provide a non-trivial check of the AdS$_4$/CFT$_3$ correspondence
recently proposed in \cite{Assel:2011xz} by verifying the GKPW relation
in the large $N$ limit.
The CFT free energy is obtained from the previous works
\cite{Benvenuti:2011ga,Nishioka:2011dq}
on the $S^3$ partition function for
3-dimensional $\scN=4$
SCFT $T[SU(N)]$.
This is matched with the computation of the type IIB action on the corresponding gravity background.
We unexpectedly find that the leading behavior of the free energy at
 large $N$ is  $\half N^2 \ln N$. We also extend our results to richer
 $T^{\hat \rho}_{\rho}[SU(N)]$ theories and argue that $\half N^2 \ln N$ is the maximal free energy at large $N$ in this class of gauge theories.

\vspace{3cm}

\end{titlepage}

\newpage
\setcounter{page}{1}

\tableofcontents

\section{Introduction}

In this paper we study
a class of 3d $\scN=4$ SCFT
$T^{\rho}_{\hat{\rho}}[SU(N)]$
introduced in \cite{Gaiotto:2008ak},
where $\rho, \hat{\rho}$ are partitions of $N$.
This theory is a 1/2 BPS domain wall theory
inside the 4d $\scN=4$ $SU(N)$ Yang-Mills theory, and plays crucial roles in the
generalizations \cite{Drukker:2010jp,Hosomichi:2010vh} of the AGT
correspondence, as well as the connection with the 3d $SL(2)$
Chern-Simons theory \cite{Terashima:2011qi,Terashima:2011xe,Dimofte:2011jd}.
The $T_{\hat{\rho}}^{\rho}[SU(N)]$ theories also appear as the basic
building blocks
for the 3d mirror of the 4d $\scN=2$ Gaiotto theories \cite{Gaiotto:2009we}
compactified on $S^1$ \cite{Benini:2010uu,Nishioka:2011dq}.

The type IIB supergravity dual for $T^{\rho}_{\hat{\rho}}[SU(N)]$
theories has recently been constructed in \cite{Assel:2011xz}.
In this paper we provide further quantitative consistency checks of this AdS4/CFT3
correspondence by verifying the GKPW relation \cite{Witten:1998qj,Gubser:1998bc} in the leading large $N$
limit.

On the CFT side, we take the large $N$ limit of the $S^3$ partition
functions of \cite{Benvenuti:2011ga,Nishioka:2011dq}, evaluated at the conformal point.
On the gravity side, we evaluate the gravity action in the gravity
background of \cite{Assel:2011xz}.
We find that in both cases the leading contribution of the free energy in the large $N$
 limit scales as
\begin{equation*}
F\sim N^2 \ln N+{\cal O}(N^2) \ .
\end{equation*}
More detailed statements will be given momentarily
in section \ref{subsec.summary}.
As we will see,
on the CFT side $N^2 \ln N$ comes from the asymptotic behavior of the
Barnes $G$-function.
On the gravity side, a factor of $N^2$ comes from the
local scaling of the supergravity Lagrangian, and an extra $\ln N$
comes from the size of the
geometry.

The organization of this paper is as follows.
We first summarize the notations and the main results (section
\ref{sec.summary}).
We then give the derivations the results in gauge theory (section
\ref{sec.gauge})
and gravity (section \ref{sec.gravity}).
We also include two appendices.

\section{Summary of the Results}\label{sec.summary}

\subsection{Review of $T^{\rho}_{\hat{\rho}}[SU(N)]$ Theories}
Let us first briefly summarize the basics of $T^{\rho}_{\hat{\rho}}[SU(N)]$
theories needed for the understanding of this paper (see
\cite{Gaiotto:2008ak}
for details).

As stated in the introduction, the $T^{\rho}_{\hat{\rho}}[SU(N)]$ theory is specified
by two partitions $\rho$ and $\hat\rho$ of $N$:
\begin{align}
\begin{split}
\rho &= \Big[ \overbrace{l^{(1)},l^{(1)},..,l^{(1)}}^{N_{5}^{(1)}},\ \overbrace{l^{(2)},l^{(2)},..,l^{(2)}}^{N_{5}^{(2)}},\ ...\ ,
\ \overbrace{l^{(p)},l^{(p)},..,l^{(p)}}^{N_{5}^{(p)}} \Big] \ ,   \\
\hat \rho &=
\Big[ \overbrace{\hat l^{(1)},\hat l^{(1)},..,\hat l^{(1)}}^{\hat N_{5}^{(1)}},\ \overbrace{\hat l^{(2)},\hat l^{(2)},..,\hat l^{(2)}}^{\hat N_{5}^{(2)}},\ ...\ ,\
\overbrace{\hat l^{(\hat p)},\hat l^{(\hat p)},..,\hat l^{(\hat
 p)}}^{\hat N_{5}^{(\hat p)}} \Big] \ ,
\end{split}
\label{rhoconvention}
\end{align}
where $l^{(a-1)}>l^{(a)}, \hat{l}^{(a-1)}>\hat{l}^{(a)}$ for all $a$ and
\beq
\sum_{a=1}^p l^{(a)} N_5^{(a)}=\sum_{a=1}^{\hat{p}}\hat{l}^{(a)} \hat{N}_5^{(a)}=N \ .
\label{sumN}
\eeq
As the notation suggests, $N_5^{(a)}$ and $\hat{N}_5^{(a)}$ represent the
      5-branes charges
of the supergravity solution (see section \ref{sec.gravity}).

To construct the 3d theory, it is useful to use the brane
configurations of \cite{Hanany:1996ie}. Namely, we consider a D3-D5-NS5-brane configuration with
$N$ D3-branes suspended between NS5-branes on the left and D5-branes on the right,
where $l^{(a)}$ D3-branes ($\hat{l}^{(a)}$ D3-branes) end on the $i$-th D5-brane
(NS5-brane).
We can identify the 3d theory after suitable exchanges of D5 and
NS5-branes.
The result is a 3d $\scN=4$ quiver gauge theory.
This theory has a non-trivial irreducible IR fixed point only when
\cite{Gaiotto:2008ak}
\beq
\rho^T > \hat{\rho} \quad  \Leftrightarrow \quad \hat{\rho}^T > \rho \ ,
\label{IRcondition}
\eeq
where $\rho > \hat{\rho}$ for $\rho=[n_1, n_2, \ldots]$ and
$\hat{\rho}=[m_1, m_2,
\ldots]$
is defined by
\beq
\sum_{i=1}^k n_i > \sum_{i=1}^k m_i
\eeq
for all $k$.  When the inequality is saturated for some value of $i$, the quiver breaks into pieces and the IR fixed point consists of products of irreducible theories.

The global symmetry of $T^{\hat{\rho}}_{\rho}[SU(N)]$
is given by $G_{\rho}\times G_{\hat{\rho}}$,
where $G_{\rho}$ is a subgroup of $\SU(N)$ commuting with the embedding $\rho$
\beq
G_{\rho}=S(\U(N_5^{(1)}) \times \cdots \times \U(N_5^{(p)})) \ .
\eeq
$G_{\rho}$ is a symmetry of the Lagrangian, and acts non-trivially on the
Higgs branch, whereas $G_{\hat{\rho}}$ is a
quantum mechanical symmetry acting on the Coulomb branch\footnote{The Cartan of this symmetry is the shift of the dual photon,
and is present in the Lagrangian.}.
We can weakly gauge these symmetries to introduce a set of real mass parameters
and FI parameters, which we collectively denote by $m$ and $\hat{m}$, respectively.
The two global symmetries are related by 3d mirror symmetry \cite{Intriligator:1996ex}
exchanging Higgs and
Coulomb branches, together with real mass and FI parameters.
This is simply the S-duality of the D3-D5-NS5 system, and in particular,
$T^{\rho}_{\hat{\rho}}[SU(N)]$ is the mirror of $T_{\hat{\rho}}^{\rho}[SU(N)]$.

\subsection{Large $N$ Free Energy}\label{subsec.summary}

We will verify the GKPW relation in the large $N$ limit:
\beq
Z_{\rm CFT}=e^{-S_{\rm gravity}} \ , \quad
\textrm{i.e.}
\quad
F_{\rm CFT}=S_{\rm gravity} \ ,
\eeq
where $Z_{\rm CFT}$ is a CFT partition function on $S^3$,
$F_{\rm CFT}:=-\ln Z_{\rm CFT}$ is the free energy, and
$S_{\rm gravity}$ is the action for the type IIB supergravity
holographic dual to the CFT.

Our findings are summarized as follows.

\begin{itemize}
\item The simplest prototypical example is the $T[SU(N)]$ theory, which
      is a $T^{\rho}_{\hat{\rho}}[SU(N))]$ theory with
\beq
\rho = \hat\rho = \big[\overbrace{1,1,...,1}^{N}\big] \ .
\label{rhospecial}
\eeq

In this case we find
\beq
F_{\rm CFT}=S_{\rm gravity}=\frac{1}{2} N^2 \ln N +\scO(N^2) \ .
\label{Fspecial}
\eeq

\item More generally we consider the case $\hat{p}=1$, i.e.,
\begin{align}
\begin{split}
\rho &= \Big[ \overbrace{l^{(1)},l^{(1)},..,l^{(1)}}^{N_{5}^{(1)}},\ \overbrace{l^{(2)},l^{(2)},..,l^{(2)}}^{N_{5}^{(2)}},\ ...\ ,
\ \overbrace{l^{(p)},l^{(p)},..,l^{(p)}}^{N_{5}^{(p)}} \Big] \ ,   \\
\hat \rho &=
\Big[ \overbrace{\hat l,\hat l,..,\hat l}^{\hat N_5} \Big] \ .
\end{split}
\label{rhogeneral}
\end{align}
We take the scaling limit
\beq
N_5^{(a)} =N^{1-\kappa_a} \gamma_a,\quad l^{(a)} = N^{\kappa_a}
\lambda^{(a)},\quad
\hat N_5 = N \hat \gamma \ ,
\label{scaling}
\eeq
where we take $N$ large, while keeping $\kappa_a, \lambda^{(a)}, \gamma_a, \hat{\gamma}$
finite.
We require
\beq
\kappa_{a-1}\ge \kappa_a,\quad  0\le \kappa_a<1,   \quad \textrm{for all  } a
\ .
\label{kappa}
\eeq
The first condition is necessary for $\rho$ to be partition, and the
      second ensures that the $N_5^{(i)}$ becomes large, hence
      justifying the validity of the supergravity solution.
We also have from \eqref{sumN} the constraint
\beq
\sum_{a=1}^p \gamma_a \lambda^{(a)}=\hat{\gamma}\, \hat{l}=1 \ .
\label{sumN2}
\eeq
In this more general case we find (CFT analysis will be provided for $\hat{l}=1$,
      and gravity analysis for general $\hat{l}$):
\begin{align}
F_{\rm CFT}=S_{\rm gravity}= \frac{1}{2} N^2\ln N \left[  (1-\kappa_1)
+ \sum_{i=2}^p \left( \sum_{a=i}^p \gamma_a \lambda^{(a)} \right)^2\left( \kappa_{i-1}-\kappa_i\right)
\right] + {\cal O}(N^2).
\label{Fgeneral}
\end{align}
In particular when all $\kappa_a=0$, i.e. when all $l^{(a)}$ are finite,
      the leading large $N$ behavior
      coincides with that in \eqref{Fspecial}.
\end{itemize}

Note the number inside the bracket in \eqref{Fgeneral} is a non-negative number smaller than $1$ due to \eqref{kappa}.
Motivated by this result we conjecture
\beq
F_{T^{\rho}_{\hat{\rho}}[SU(N)]}\le F_{T[SU(N)]} \ .
\label{conj}
\eeq
for all $\rho, \hat{\rho}$ satisfying \eqref{IRcondition}.
It would be interesting to see if some of the above inequalities could be explained in terms of the F-theorem \cite{Jafferis:2010un,Jafferis:2011zi}
and the RG flows between the fixed points.
The rest of this paper will be devoted to the derivation of \eqref{Fspecial} and \eqref{Fgeneral}.

\section{CFT Analysis}\label{sec.gauge}

In this section we analyze the CFT free energy $F_{\rm CFT}$.

\subsection{The $S^3$ Partition Function}

Let us begin with the $T[SU(N)]$ theory \eqref{rhospecial}.
The partition function of this theory was computed by localization \cite{Kapustin:2009kz}
to be
\cite{Benvenuti:2011ga,Nishioka:2011dq} (see also
\cite{Gulotta:2011si}):
\footnote{
The expression in \cite{Nishioka:2011dq} contains an extra factor of
$1/N!$.
However, this factor does not alter the leading behavior of the free
energy and hence will be dropped in this paper.
}
\beq
Z_{S^3}[T[SU(N)]](m, \hat{m})=\frac{\sum_{w\in \mathfrak{S}_N}
(-1)^w e^{2\pi \imath m\cdot w(\hat{m})}}{\Delta(m) \Delta(\hat{m})},
\label{ZTSUN}
\eeq
where $(-1)^w$ is a sign of a permutation $w\in\mathfrak{S}_N$, $m=(m_1, \ldots, m_N)$
with $\sum_i m_i=0$
($\hat{m}=(\hat{m}_1, \ldots, \hat{m}_N)$ with $\sum_i \hat{m}_i=0$) are the FI
parameters\footnote{FI parameters are actually the differences of
$\hat{m}_i$. However, we will loosely refer to $\hat{m}$ as FI
parameters.} (real mass parameters), and
$$
m\cdot w(\hat{m}):=\sum_i
m_i \, \hat{m}_{w(i)} \ ,
$$
and $\Delta$ is the (sinh) Vandermonde determinant
\beq
\Delta(m):=\prod_{i<j} 2 \sinh \pi (m_i-m_j), \quad
\Delta(\hat{m}):=\prod_{i<j} 2 \sinh \pi (\hat{m}_i-\hat{m}_j). \
\eeq

\begin{figure}[htbp]
\vspace{-1.7cm}
\centering
\includegraphics[width=0.5\textwidth]{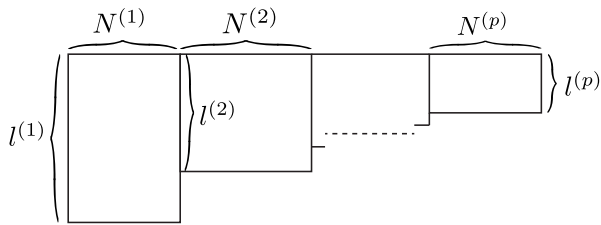}
\vspace{-1.7cm}
\caption{We decompose the young diagram corresponding to $\rho$ into $p$
 blocks, see \eqref{rhoconvention}.}
\label{fig.rule}
\end{figure}

For more general $T^{\hat{\rho}}_{\rho}[SU(N)]$ theories,
the partition function takes a similar form as in \eqref{ZTSUN} \cite{Nishioka:2011dq}:
\beq
Z_{S^3}[T^{\rho}_{\hat{\rho}}[SU(N)]](m, \hat{m})=\frac{\sum_{w\in \mathfrak{S}_N}
(-1)^w e^{2\pi \imath m_{\rho}\cdot w(\hat{m}_{\hat{\rho}})}}{\Delta_{\rho}(m) \Delta_{\hat{\rho}}(\hat{m})}.
\label{ZTSUN2}
\eeq
Here $m_{\rho}, \hat{m}_{\hat{\rho}}$ are $N$-vectors, and each of their
 components
is associated with a box of the Young diagram (also denoted by $\rho, \hat{\rho}$)
corresponding to the partitions $\rho, \hat{\rho}$.
For later purposes let us describe them by
dividing the boxes of $\rho$ into $p$
blocks, where the $a$-th block is a rectangle with rows of length $N_5^{(a)}$
and columns of length $l^{(a)}$ (recall \eqref{rhoconvention}, and
see fig. \ref{fig.rule}). A box of
$\rho$ could then be labeled by
a triple $(a, i, \alpha)$ with
$1\le a\le p, 1\le i \le N_5^{(a)}, 1\le \alpha
\le l^{(a)}$, where $a$ is the label for the block
and $i$ ($\alpha$) is the label for the column (row) inside the $a$-th block.
The same applies to $\hat{\rho}$.
In this notation, we have
\beq
(m_{\rho})_{(a,i, \alpha)}=\imath(w_{l^{(a)}})_{\alpha}+ m_{a,i} , \quad
(\hat{m}_{\hat{\rho}})_{(a,i, \alpha)}=\imath(w_{\hat{l}^{(a)}}
)_{\alpha}+ \hat{m}_{a,i} \ ,
\eeq
where $w_N$ is a Weyl vector of the $SU(N)$ Lie algebra defined by
\bea
w_N=\left(\frac{N-1}{2}, \frac{N-3}{2},\ldots,
-\frac{N-1}{2}\right)\ .
\label{rhoW}
\eea
Also, $\Delta_{\rho}(m)$ and $\Delta_{\hat{\rho}}(\hat{m})$ are defined by
\begin{equation}
\begin{split}
\Delta_{\rho}(m)& =\prod_p \prod_{q<r} 2 \sinh \pi ((m_{\rho})_{[p,q
]}-(m_{\rho})_{[p,r]}) ,
\\
\Delta_{\hat{\rho}}(\hat{m})& =\prod_p \prod_{q<r} 2 \sinh \pi
((\hat{m}_{\hat{\rho}})_{[p,q]}-(\hat{m}_{\hat \rho})_{[p,r]}) \ ,
\end{split}
 \end{equation}
where $[p,q]$ represents a box inside $\rho, \hat{\rho}$ at row $p$ and column $q$.
Note that the $(m_{\rho})_{[p,q]}$ are simply a relabeling of the $(m_{\rho})_{(a,i,\alpha)}$ introduced previously.

Several remarks are now in order.
First, the partition function \eqref{ZTSUN2} is manifestly invariant
under
the simultaneous exchange of $\rho, \hat{\rho}$ and $m, \hat{m}$.
This is a manifestation of the 3d mirror
symmetry.

Second, \eqref{ZTSUN2} vanishes unless $\rho^T\ge \hat{\rho}$ \cite{Nishioka:2011dq}. This is
consistent with the condition \eqref{IRcondition}
for the existence of a non-trivial IR SCFT. This condition has a
counterpart in the gravity dual \cite{Assel:2011xz}.

Third, the expression \eqref{ZTSUN} is either
real or pure imaginary, however there is an ambiguity of the phase of the
$S^3$ partition function and we will hereafter concentrate on the
absolute value of the $S^3$ partition function.

\subsection{$T[SU(N)]$}
\label{subsec.largeN}

Let us study the large $N$ behavior of our partition functions.

For clarity, let us begin with the $T[SU(N)]$ theories,
whose partition function is given in \eqref{ZTSUN}.
When the parameters $m$ and $\hat{m}$ are
generic and kept finite in the limit,\footnote{By generic we mean that there are no cancellations in the sum
in the numerator of
\eqref{ZTSUN}.}
we have $\sum_{w\in \mathfrak{S}_N} \sim \scO(N!)$, whose
logarithm contributes $\scO(N \ln N)$ to the $F_{\rm CFT}$.
The remaining contributions come from the two sinh Vandermonde determinants,
each of which involves ${N\choose 2}\sim \scO(N^2)$ terms.
This gives
\beq
F_{\rm CFT}\sim \mathcal{O}(N^2) \ .
\eeq
This is not surprising since after all our theories are standard gauge
theories.

However, the scaling behavior could change if we consider non-generic values of
$m$ and $\hat{m}$.
This is exactly happens to our CFT case, where we need to take the limit
$m, \hat{m}\to 0$ of \eqref{ZTSUN}:
\beq
Z_{\rm CFT}=\lim_{m, \hat{m}\to 0} \big| Z_{S^3} \big| \ .
\eeq

We choose to take the limit in two steps.
First, let us take the $\hat{m}\to 0$ limit of \eqref{ZTSUN2} with $\hat{\rho}=[1,\ldots,
1]$.
This is conveniently done by setting
$\hat{m}=\epsilon w_N$ and by taking $\epsilon\to 0$,
where $w_N$ is defined in \eqref{rhoW}.
Using the Weyl denominator formula, we have
\bea
\sum_{w\in \mathfrak{S}_N} (-1)^w e^{2\pi \imath \epsilon w_N \cdot
m_{\rho}}=\prod_{\alpha>0} 2\imath \sin\left(\pi \epsilon \alpha\cdot
m_{\rho}\right)=\prod_{j<k} 2\imath \sin\left(\pi \epsilon
(m_j-m_k)\right) \ .
\eea
In the limit $\epsilon_2\to 0$, this cancels the factor
$\Delta(\epsilon w_N)=\prod_{j<k} 2\sinh \pi (\epsilon (j-k))$, giving
\beq
 \left| \prod_{j<k} \frac{\imath}{(j-k)} \frac{\prod_{j<k}
 (\unhat{m}_{\rho})_j-(\unhat{m}_{\rho})_k}{\Delta_{\rho}(\unhat{m})}\right|
=
 \frac{1}{G_2(N+1)}
\left|
 \frac{\prod_{j<k}
 (\unhat{m}_{\rho})_j-(\unhat{m}_{\rho})_k}{\Delta_{\rho}(\unhat{m})}
\right| \ .
\label{halflimit}
\eeq

We next need to take the limit $\unhat{m}\to 0$.
This is easy for our case, $\rho=[1,\ldots, 1]$;
\[
  \frac{\prod_{i<j}
 (\unhat{m}_{\rho})_i-(\unhat{m}_{\rho})_j}{\Delta_{\rho}(\unhat{m})}=
  \prod_{i<j}
 \frac{\unhat{m}_i-\unhat{m}_j}{2 \sinh \pi (\unhat{m}_i-\unhat{m}_j)}\to (2\pi)^{-\frac{N(N-1)}{2}},
\]
which gives
\beq
Z_{\rm CFT}=\frac{1}{(N-1)!(N-2)! \ldots 2!
1!}\left(\frac{1}{2\pi}\right)^{\frac{N(N-1)}{2}} =\frac{1}{G(N+1)} \left(\frac{1}{2\pi}\right)^{\frac{N(N-1)}{2}},
\label{ZTSUNzero}
\eeq
where $G_2(x)$ is the Barnes $G$-function defined in Appendix \ref{app.Barnes}.
From the asymptotics of $G_2(x)$ \eqref{Gasymptotics}, we have
\beq
F_{\rm CFT}=\frac{N^2}{2} \ln N +\left[-\frac{3}{4}-\frac{1}{2} \ln
\left(\frac{1}{2\pi} \right) \right] N^2 + \mathcal{O}(N \ln
N) \ ,
\label{cftparttsun}
\eeq
which gives \eqref{Fspecial}.

\subsection{$T^{\rho}_{\hat{\rho}}[SU(N)]$}

Let us consider the more general case given in \eqref{rhogeneral}.

As long as $\hat{\rho}=[1,\ldots, 1]$
the argument of the previous subsection works
up until \eqref{halflimit}.
In \eqref{halflimit} we already have a factor of $G_2(N+1)$.
Just as in the $T[SU(N)]$ case, this contributes
\begin{equation}
\frac{1}{2}N^2 \ln N \ ,
\label{part1}
\end{equation}
to the free energy.
Next, let us send the FI parameters to zero in \eqref{halflimit}.
The denominator $\Delta_{\rho}(\unhat{m})$ goes to zero in the limit,
but the same is true for the numerators, yielding the finite answer.
We obtain powers of $2\pi$ in this process from the limit of
$\Delta_{\rho}(\unhat{m})$,
however this only gives a subleading contribution of order $N^2$.

There are still contributions from the numerator $\prod_{i<j}
\left[(\unhat{m}_{\rho})_i-(\unhat{m}_{\rho})_j\right]$, which we have not yet
taken into account. In the notation of the previous
section
the limit of this contribution is\footnote{A small modification is needed for the formula \eqref{alphabeta} when $\alpha-\beta$ is odd. This does not, however,
affect the leading behavior of the free energy.
}
\begin{equation*}
(\unhat{m}_{\rho})_{(a,i,\alpha)}-(\unhat{m}_{\rho})_{(b,j,\beta)}
=\imath\left[(w_{l^{(a)}})_{\alpha}-(w_{l^{(b)}})_{\beta}\right]=\imath(\alpha-\beta) \ ,
\label{alphabeta}
\end{equation*}
where $1\le \alpha\le l^{(a)}, 1\le \beta\le l^{(b)}$ and $\alpha \ne \beta$.

When the two boxes are in the same block,
this contributes a factor
\begin{align*}
\left(N_5^{(a)}\right)^2 \ln \left[ (l^{(a)}-1)! (l^{(a)}-2)! \ldots 1!
			     \right] \ ,
\end{align*}
where the factor $\left(N_5^{(a)}\right)^2$ accounts for the degeneracy from the
column labels $i$.
This contributes, under the scaling \eqref{scaling},
\begin{align}
-\frac{1}{2} \left[\kappa_a
 (\lambda^{(a)}\gamma_a)^2\right]N^2 \ln N  +\mathcal{O}(N^2)
\ ,
\label{part2}
\end{align}
to the free energy.
When the two boxes are in the different blocks $a, b$
with $l^{(a)}\ge l^{(b)}, \kappa_a\ge \kappa_b$,
the contribution to the free energy is
\begin{align*}
-2\left(N_5^{(a)} N_5^{(b)}\right) \ln \left[
 \left(\frac{l^{(a)}+l^{(b)}}{2}-1\right)!
 \left(\frac{l^{(a)}+l^{(b)}}{2}-2\right)!  \ldots \left(\frac{ l^{(a)}-l^{(b)}}{2}\right)!
			     \right]  \ .
\end{align*}
The expression inside the bracket gives
\begin{equation*}
\ln \left[ G_2\left(\frac{l^{(a)}+l^{(b)}}{2}+1\right)\right]
-\ln
 \left[G_2\left(\frac{l^{(a)}-l^{(b)}}{2}+1\right)\right]
\sim \frac{1}{2}l^{(a)} l^{(b)} \ln l^{(a)} \ .
\end{equation*}
Thus the contribution amounts to
\begin{align}
-2\left(N_5^{(a)} N_5^{(b)}\right)
\frac{1}{2}l^{(a)} l^{(b)} \ln l^{(a)} =-2\frac{1}{2}\left[(\lambda^{(a)}
 \gamma_a \lambda^{(b)} \gamma_b) \kappa_a\right]
 N^2 \ln N \ .
\label{part3}
\end{align}
Collecting all the contributions \eqref{part1}, \eqref{part2} and \eqref{part3}, we have
\beq
F_{\rm CFT}=\frac{1}{2} N^2 \ln N\left[1-\sum_{a=1}^p
(\lambda^{(a)} \gamma_a)^2 \kappa_a -2\sum_{a\ne b,\, l^{(a)}>l^{(b)}}
(\lambda^{(a)} \gamma_a \lambda^{(b)} \gamma_b) \kappa_a \right]
+\scO(N^2) \ .
\eeq
From \eqref{sumN2} we can show that this coincides with \eqref{Fgeneral}.


In all of the examples above, the leading contribution to the partition
function comes from the Barnes $G$-functions. It is curious to note that
the same function appears in the formula for the volumes of Lie group $SU(N)$
\cite{Macdonald}, and hence in the measure for the $SU(N)$ gauge theory.
This is probably not a coincidence, since
in the correspondence in \cite{Nishioka:2011dq} the $S^3$
partition function
of $T^{\rho}_{\hat{\rho}}[SU(N)]$
theory is identified with an overlap of wavefunctions of a 1d quantum
mechanics, which is obtained from a dimensional reduction of
the 2d Yang-Mills theory. The measure of 2d Yang-Mills contains a
volume factor for the gauge group $U(N)$.
The same $N^2 \ln N$ type behavior appears in a number of different
contexts, such as Gaussian matrix models, $c=1$, topological string on
the conifold or more recently in the weak coupling expansion of the
ABJM theory \cite{Drukker:2010nc}.

\section{Gravity Analysis}\label{sec.gravity}

In this section we analyze the type IIB supergravity action $S_{\rm gravity}$ in the holographic dual.

\subsection{Summary of the Gravity Solution}\label{sec:solution}

First we summarize the holographic duals of the $T^{\hat{\rho}}_{\rho}[\SU(N)]$ theories constructed in
\cite{Assel:2011xz} (see also \cite{Aharony:2011yc} for related work), which is based on earlier solutions found in \cite{D'Hoker:2007xy,D'Hoker:2007xz}.

The geometry of the type IIB backgrounds is an AdS$_4\times S^2\times S^2$ fibration over a two-dimensional Riemann surface $\Sigma$.  We will parameterize $\Sigma$ by an infinite strip, although it will turn out that $\Sigma$ has finite volume and is really compact.  Next we introduce complex coordinates on $\Sigma$ as $z, \bar{z}$.  We will also make use of the real coordinates defined by writing $z=x+ \imath y$.  After fixing $\Sigma$, the solution is then determined by two real harmonic functions, $h_1$ and $h_2$, on $\Sigma$.

The metric can be written as
\begin{align}
ds^2 = f_4^2 ds^2_{AdS_4} + f_1^2 ds^2_{S^2_1} + f_2^2 ds^2_{S^2_2} + 4 \rho^2 dz d\bar z\ ,
\end{align}
where the warp factors are given by
\begin{align}
f_4^8 = 16 \frac{N_1 N_2}{W^2} \ ,
\ \ \
f_{1}^8 = 16 h_{1}^8 \frac{N_{2} W^2}{N_{1}^3} \ ,
\ \ \
f_{2}^8 = 16 h_{2}^8 \frac{N_{1} W^2}{N_{2}^3} \ ,
\ \ \
\rho^8 = \frac{N_1 N_2 W^2}{h_1^4 h_2^4} \ ,
\end{align}
and we defined the auxiliary functions
\begin{align}\label{W}
W = \p \bar \p(h_1 h_2) \ , \qquad N_{j} = 2 h_1 h_2 |\p h_{j}|^2 -
 h_{j}^2 W \ .
\end{align}
This geometry is supported by non-vanishing ``matter'' fields, which include the dilaton field
\begin{align}
e^{2 \phi} = \sqrt{N_2 \over  N_1 } \ ,
\end{align}
in addition to non-vanishing 3-form and 5-form fluxes which are given in appendix \ref{app.Fluxes}.

\begin{figure}
\vspace{-2cm}
\centering
\includegraphics[width=12cm]{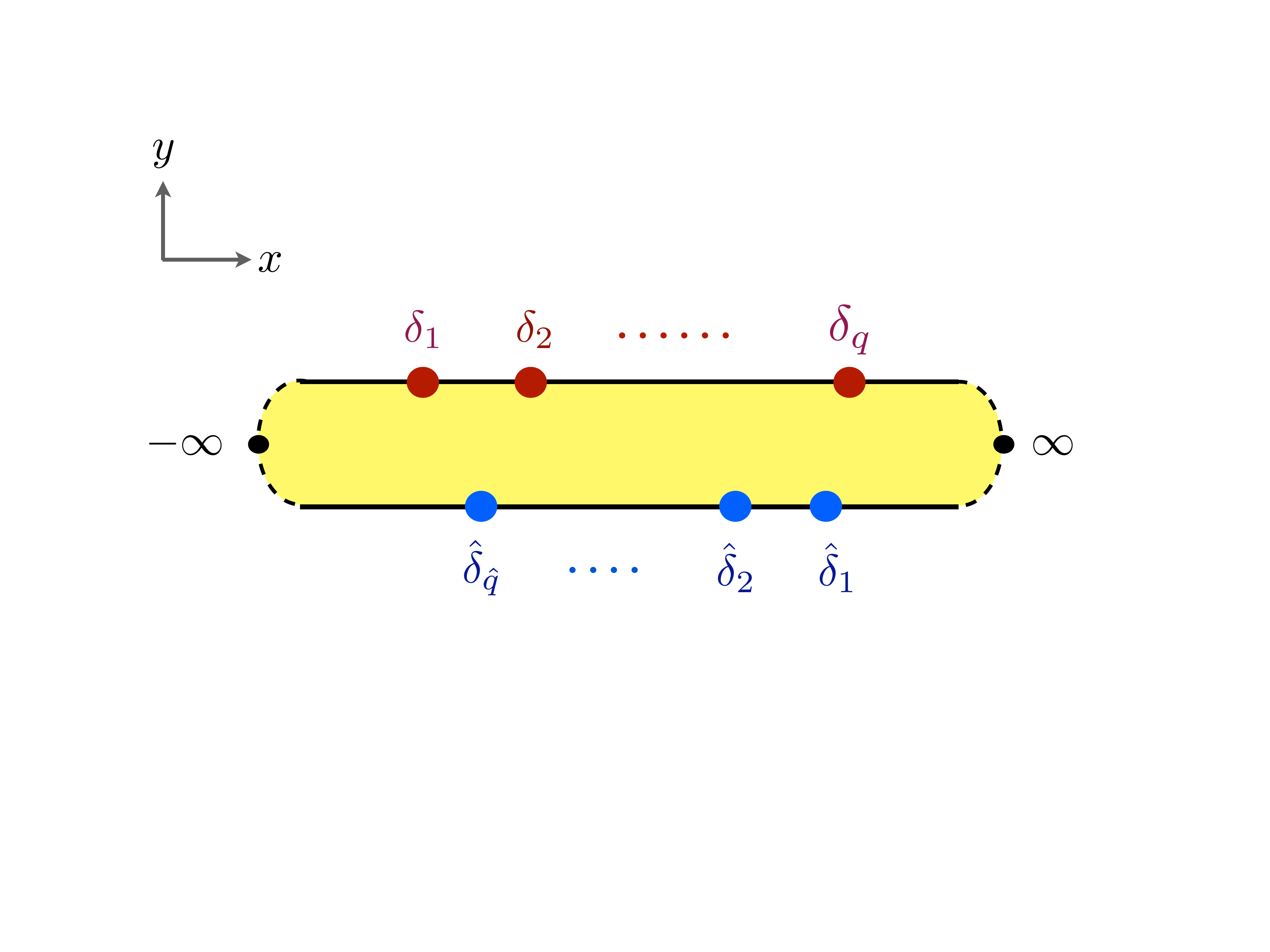}
\vspace{-3cm}
\caption{The infinite strip with logarithmic singularities corresponding to stacks of five-branes.  The upper (red) singularities correspond to D5-branes, while the lower (blue) singularities correspond to NS5-branes. The geometry smoothly caps off into an $S^6$ as $x \rightarrow \pm \infty$.}
\label{sugrafig}
\end{figure}

We now turn to the specific solutions corresponding to
$T^{\hat{\rho}}_{\rho}[SU(N)]$.
The classical supergravity solutions describing the near horizon limit of D3-branes suspended between $p$ stacks of D5-branes and $\hat p$ stacks of NS5-branes is given by the two harmonic functions:
\begin{align}
\label{harm1}
\begin{split}
h_1 & =  -  \sum_{a=1}^{p} \frac{\alpha'}{4} N_5^{(a)} \ln\left[ \tanh\left(
{\small  {\frac{\imath \pi}{4}-\frac{z-\delta_a}{2}
}}
\right)\right] + c.c. \ , \\
h_2 &=  - \sum_{b=1}^{\hat{p}} \frac{\alpha'}{4} \hat N_5^{(b)} \ln\left[ \tanh\left(\frac{ z-\hat{\delta}_{b} }{2}\right)\right] + c.c.    \ ,
\end{split}
\end{align}
with $- \infty < x < \infty$ and $0 \leq y \leq \pi/2$.  Here $\delta_1 < \delta_2 < ... < \delta_p$ are the positions of D5-brane singularities on the upper boundary of the strip ($y = \pi/2$), whereas $\hat \delta_1 > \hat \delta_2 > ... > \hat \delta_{\hat p}$ are the positions of NS5-brane singularities on the lower boundary ($y = 0$) (see fig. \ref{sugrafig}).  The points at $x = \pm \infty$ are regular interior points of the ten-dimensional geometry.

The coefficients of the logarithms determine the number of 5-branes located at the singularities.  The number of D5-branes located at $\delta_a$ is denoted by $N_5^{(a)}$, while the number of NS5-branes located at $\hat \delta_{(b)}$ is denoted by $\hat N_5^{(b)}$.  Unbroken supersymmetry requires that there are only branes (or only anti-branes) of each kind. Thus all the $N_5^{(a)}$ must have the same sign, and likewise for all the $\hat N_5^{(b)}$.  This positivity condition is also necessary for smoothness away from the 5-brane singularities.  The net number of D3-branes ending on the $a$-th D5-brane stack is denoted by $N_3^{(a)}$, while the number of D3-branes ending on the $b$-th NS5-brane stack is denoted by $\hat N_3^{(b)}$.  These quantities are determined by the locations $\delta_a$ and $\hat \delta_b$, of the 5-brane stacks as
\begin{align}
\begin{split}
&N_3^{(a)} =  N_5^{(a)} \sum_{b=1}^{\hat p} \hat N_5^{(b)} \, {2\over\pi} {\rm arctan}(e^{\hat\delta_b -\delta_a}) \ ,
\\
&\hat N_3^{(b)} = \hat N_5^{(b)} \sum_{a=1}^{p} N_5^{(a)} \, {2\over\pi} {\rm arctan}(e^{\hat\delta_b -\delta_a})\ .
\end{split}
\label{ginvN3}
\end{align}
We define the total number of D3-branes as $N \equiv \sum_{a=1}^p
 N_3^{(a)} =  \sum_{b=1}^{\hat p} \hat N_3^{(b)}$,
and the linking numbers by $l^{(a)} = N_3^{(a)}/N_5^{(a)}$  and \,   $\hat l^{(b)} = \hat N_3^{(b)}/\hat N_5^{(b)}$.\footnote{The relations
between the integer brane charges and the supergravity parameters are not easily inverted. To express the latter in terms
of the brane charges one must solve a system of  transcendental
equations.}  These parameters, $N, N_5^{(a)}, \hat{N}_5^{(a)}, l^{(a)},
 \hat{l}^{(a)}$, are identified with the same parameters of the same
 names in section \ref{sec.summary} under the holographic duality.

\subsection{The Gravity Action}\label{sec:action}

The type IIB action in Einstein frame is\footnote{We use the convention $|F_{(a)}|^2 = { 1 \over a!} F_{(a) \, M_1 M_2 .. M_a} F_{(a)}^{\, M_1 M_2 .. M_a}$.}
\begin{align}
\label{IIBaction}
\begin{split}
S_{\rm IIB} &=
-{1 \over 2 \kappa_{10}^2} \int d^{10}x \sqrt{g} \bigg \{
R - {4 \over 2} \p_M \phi \p^M \phi
- \half e^{4 \phi} \p_M \chi \p ^M \chi
- { 1 \over 2} e^{-2 \phi} |H_{(3)}|^2
\\& \qquad
- {1 \over 2} e^{2 \phi} |F_{(3)} + \chi H_{(3)}|^2
- {1  \over 4} | \hat F_{(5)}|^2 \bigg \}
+ {1 \over 4 \kappa _{10}^2} \int d^{10}x \ C_{(4)} \wedge H_{(3)} \wedge F_{(3)} \ ,
\end{split}
\end{align}
where one imposes the self-duality condition $\hat F_{(5)} = * F_{(5)}$ as a supplementary equation.  The coupling $\kappa_{10}$ is related to the string scale $\alpha'$ by $2\kappa_{10}^2 = (2\pi)^7 (\alpha')^4$.

Due to the presence of the self-duality condition, the action \eqref{IIBaction} cannot be directly used to compute the on-shell value of the action.  One way to deal with this is to relax the requirement of Lorentz invariance of the action.  In this case an action principle could be obtained along the lines of \cite{Henneaux:1988gg}.  As suggested in \cite{Giddings:2001yu}, perhaps the easiest way to implement this for the full type IIB supergravity action is to make a T-duality transformation of the type IIA action.  A simpler method is to first dimensionally reduce the theory to 4-dimensions.  After carrying out the dimensional reduction, one can then truncate the theory to the 4-dimensional graviton.  To see this is consistent, one may check that the solutions of \cite{D'Hoker:2007xy,D'Hoker:2007xz} can be extended by replacing the AdS$_4$ space with any space which obeys the same Einstein equations.  Thus truncating to the 4-dimensional graviton is a consistent truncation.\footnote{To see this more explicitly, first consider the 10-dimensional metric $ds^2 = f_4^2 ds_{(4)}^2 + f_1^2 ds^2_{S_1^2} + f_2^2 ds^2_{S_2^2} + 4 \rho dz d\bar z^2$, where $ds_{(4)}^2$ is an arbitrary 4-dimensional metric.  This is a solution to the type IIB supergravity equations of motion as long as the 4-dimensional Ricci tensor satisfies $R_{(4) \mu\nu} = -3 g_{(4)\mu\nu}$.  One can then write the 10-dimensional Ricci scalar as $R = f_4^{-2} R_{(4)} + ...$, where the omitted terms do not depend on $ds_{(4)}^2$.  The action then takes the form $S =- \frac{1}{2 \kappa_{10}^2} \int d^{10}x (f_4 f_1 f_2)^2 4 \rho^2 \sqrt{g_{(4)}} (R_{(4)} + ...)$, where again the omitted terms do not depend on $ds_{(4)}^2$.  Requiring the variation with respect to $ds_{(4)}^2$ to now reproduce the correct equation of motion yields the effective action \eqref{acteff0}.}

The effective action for this mode is given by
\begin{align}
\label{acteff0}
S_{\rm eff} =- \frac{1}{2 \kappa_{10}^2} {\rm vol}_6
 \int_{\textrm{AdS}_4} \! d^4x \sqrt{g_{(4)}} (R_{(4)} + 6) \ ,
\end{align}
where the cosmological constant has been chosen so that the unit AdS$_4$ space is a solution.  The subscript $(4)$ reminds us that $g_{(4)}$ is the 4-dimensional metric and $R_{(4)}$ is the associated Ricci scalar.  The quantity ${\rm vol}_6$ follows from the initial dimensional reduction and is the volume of the internal space dressed appropriately with the warp factor of AdS$_4$
\begin{align}
{\rm vol}_6 = (4 \pi)^2 \int_{\Sigma} d^2x (f_4 f_1 f_2)^2 4 \rho^2
= 32 (4 \pi)^2 \int_{\Sigma} d^2x (-W) h_1 h_2 \ .
\label{vol6}
\end{align}
The specific solution we are interested in is AdS$_4$ with Ricci scalar $R_{(4)} = -12$.  Thus the on-shell action becomes simply
\begin{align}
\label{acteff}
S_{\rm eff} = -\frac{1}{(2\pi)^7 (\alpha')^4} {\rm vol}_6 \left( \frac{4}{3} \pi^2 \right) (-6) \ ,
\end{align}
where we have used the regularized volume of AdS$_4$, ${\rm vol}_{AdS_4} = (4/3) \pi^2$, which may be computed using holographic regularization \cite{Henningson:1998gx,Balasubramanian:1999re,Emparan:1999pm,de Haro:2000xn} (see for example section 5 of \cite{Marino:2011nm}).

\subsection{$T[SU(N)]$}

Let us first consider the gravity dual for $T[SU(N)]$.  The harmonic
functions are:
\begin{align}
\label{exactTSUN}
\begin{split}
h_1 &= - \frac{\alpha' N}{4} \ln \bigg[ \tanh \bigg( \frac{\imath \pi}{4} - \frac{z - \delta}{2} \bigg) \bigg] + c.c. \ , \\
h_2 &= - \frac{\alpha' N}{4} \ln \bigg[ \tanh \bigg( \frac{z + \delta}{2} \bigg) \bigg] + c.c. \ ,
\end{split}
\end{align}
with
\beq
\delta = - \half \ln\left[\tan\left(\frac{\pi}{2N}\right)\right] \ ,
\eeq
where we have used a translation to set $\hat \delta = - \delta$.  There is one stack of $N$ D5-branes at the position $z = \imath \frac{\pi}{2} - \half \ln[\tan(\frac{\pi}{2N})]$ and one stack of $N$ NS5-branes at $z = \half \ln[\tan(\frac{\pi}{2N})]$ with $N$ D3-branes stretched between them.

\begin{figure}
\centering
\includegraphics
[width=8cm]
{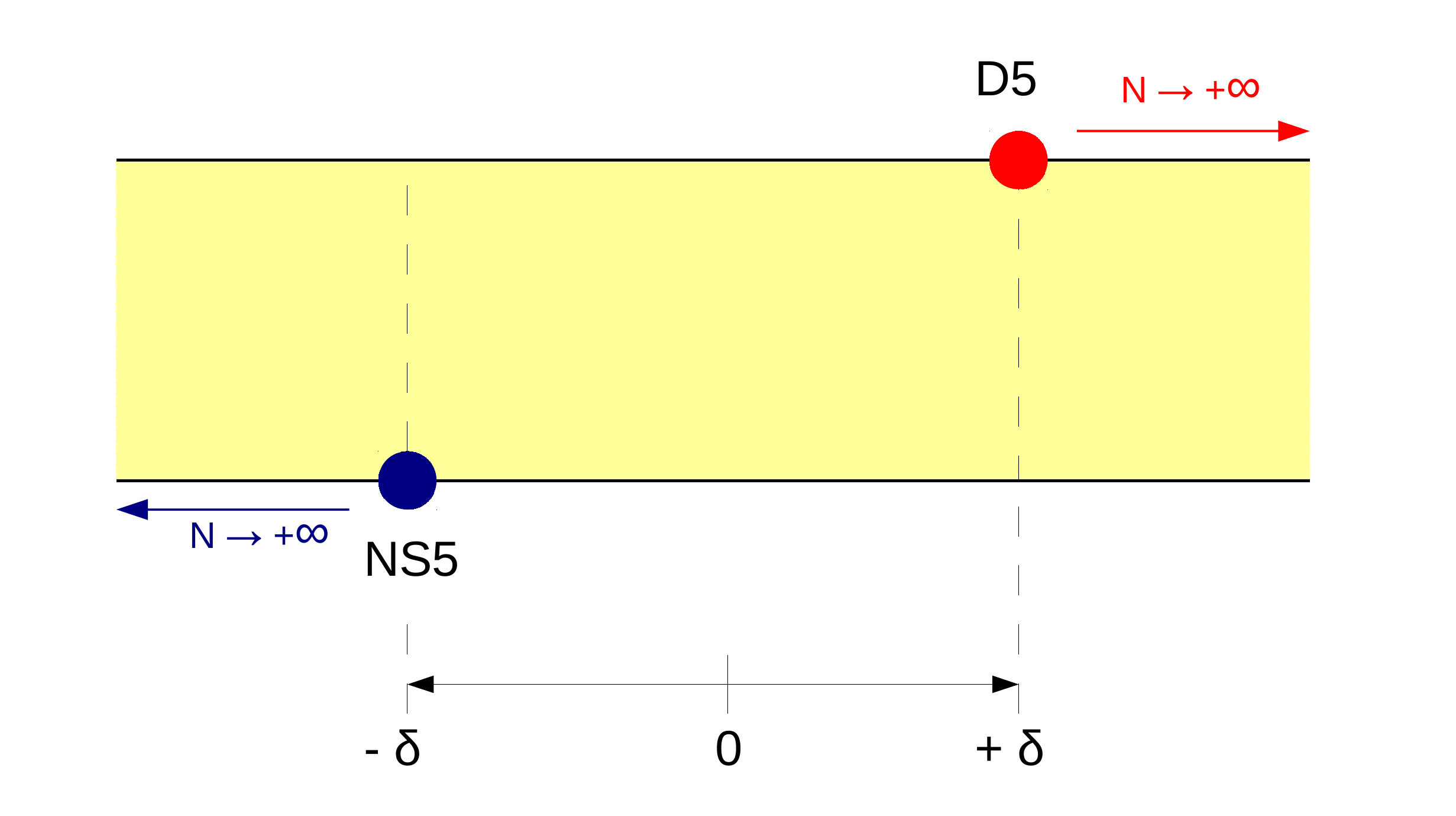}
\caption{Geometry of the $T[SU(N)]$ dual background represented by the
 strip with two 5-brane singularities at positions $\pm \delta \sim
 \pm \half \log N$. In the large $N$ limit the stacks go to $\pm$
 infinity.}
\label{fig1}
\end{figure}

We now wish to take the large $N$ limit of this configuration.  It will turn out that locally the Lagrangian density will scale with a factor of $N^2$ at leading order in $N$.  Secondly, as $N$ goes to infinity, the positions $\delta$ of the 5-brane stacks are sent to infinity in opposite directions (see fig. \ref{fig1}).  This leaves a large region of geometry between $-\delta$ and $\delta$ of size $\ln N$, which will reproduce the $\ln N$ behavior of the partition function.  Thus one can understand the leading behavior of the $T[SU(N)]$ partition function as coming from the geometry located between the two stacks of 5-branes.

To make this more explicit and also compute the exact numerical coefficient, we now work out the large $N$ expansion.  First we re-scale the $x$ coordinate so that $z = \delta x + \imath y$ and then expand the harmonic functions $h_1$ and $h_2$ around large $N$. At leading order we obtain
\begin{align}
\begin{aligned}
h_1 &= \alpha' \sin(y) N\, e^{\delta(x-1)} + ...&  &\textrm{if} \ \ x<1 \ ,\cr
    &= \alpha' \sin(y) N\, e^{\delta(1-x)} + ...&  &\textrm{if} \ \ x>1 \ ,\cr
h_2 &= \alpha' \cos(y) N\, e^{\delta(1+x)} + ...&  &\textrm{if} \ \ x<-1\ , \cr
    &= \alpha' \cos(y) N\, e^{-\delta(1+x)} + ...&  &\textrm{if} \ \ x>-1 \ .
\label{h1h2tsun}
\end{aligned}
\end{align}
From \eqref{h1h2tsun} we find that the only contribution to the action at this order comes from the central region $-1 < x < 1$.  In this region $W$ is given by $W = - \frac{1}{2} e^{-2 \delta} N^2 (\alpha')^2 \sin(2y)$.  Computing the volume of the internal space, \eqref{vol6}, and plugging into the expression for the effective action, \eqref{acteff}, we find
\begin{align}
\label{tsungravpart1}
S_{\rm eff} &=  \frac{4 N^4 \delta e^{-4 \delta}}{\pi^2} + ... \cr
&= \frac{1}{2} N^2 \ln N + {\cal O}(N^2) \ .
\end{align}
This reproduces exactly the leading order behavior of the CFT partition function \eqref{cftparttsun}.  Finally we note that including higher order terms in the expansions of the harmonic functions will give additional contributions of order $N^2$.

Since we have explicit D5-brane and NS5-brane singularities in the geometry, one may worry about the validity of our approximation.  We shall argue that the corrections due to the 5-brane singularities are at most of order $N^2$ and do not contribute to the leading $N^2 \ln N$ behavior.  To do so, we first examine the geometry in the central region in the large $N$ limit.  The metric factors are given by
\begin{align}
\begin{aligned}
f_4^2 &= \sqrt{2} \alpha' N e^{-\delta} [(2-\cos(2y))(2+\cos(2y))]^{\frac{1}{4}} \ , \cr
f_1^2 &= 2 \sqrt{2} \alpha' N e^{-\delta} \sin(y)^2  \left[ \frac{2+\cos(2y)}{(2-\cos(2y))^3} \right]^{\frac{1}{4}} \ , \cr
f_2^2 &= 2 \sqrt{2} \alpha' N e^{-\delta} \cos(y)^2  \left[  \frac{2-\cos(2y)}{(2+\cos(2y))^3} \right]^{\frac{1}{4}} \ , \cr
4 \rho^2 &= 2 \sqrt{2} \alpha' N e^{-\delta} [(2-\cos(2y))(2+\cos(2y))]^{\frac{1}{4}} \ ,
\label{tsungeom}
\end{aligned}
\end{align}
while the dilaton and fluxes are given by (see Appendix \ref{app.Fluxes})
\begin{align}
\begin{aligned}
e^\phi  &= e^{-\delta x} \left( \frac{2+ \cos(2y)}{2-\cos(2y)} \right)^{\frac{1}{4}} \ , \cr
b_1 &= 8 \alpha' N e^{-\delta(1+x)} \frac{\sin^3(y)}{2-\cos(2y)} \ , \cr
b_2 &= -8 \alpha' N e^{\delta(x-1)} \frac{\cos^3(y)}{2+\cos(2y)} \ , \cr
j_1 &= - e^{-2 \delta} N^2 (\alpha')^2 (3 x \delta + \cos(2y)) \ .
\end{aligned}
\label{tsunfluxes}
\end{align}
It is interesting to note that this is exactly the limiting geometry of Janus found in \cite{Bachas:2011xa} for the case of an infinite jump in the coupling.\footnote{The supersymmetric Janus solution is dual to ${\cal N} = 4$ super-Yang-Mills with a jumping coupling.}  The radius $L$ of the Janus space is related to $N$ by $L^2 = 2 \sqrt{2} \alpha' N e^{-\delta}$.  In the case we consider here, the $\Sigma$ space comes with a natural cutoff at $|x| = \delta$, while for Janus the space is unbounded.

We now consider curvature corrections.  Using the above formulas for the metric factor and dilaton, the string frame Ricci scalar in the central region, $-1 \leq x \leq 1$, is given by
\begin{align}
\label{riccscl}
\alpha' R &= \frac{1}{\pi 2^{1/2}} \bigg( \frac{2N}{\pi} \bigg)^{\frac{x-1}{2}} \frac{419 - 60 \cos(4y)+ \cos(8y)}{(7-\cos(4y))^2 (2+\cos(2y))^{1/2}} \ .
\end{align}
Due to the large $N$ limit, throughout most of the region we have
$\alpha' R \ll 1$.  However, due to the presence of D5-branes, as one approaches $x = 1$, $\alpha' R$ is of order one and one expects higher curvature corrections to play a role.  Since these corrections are localized only in the region near $x=1$, we expect that they do not receive the $\ln N$ enhancement and therefore contribute only at order $N^2$.  A similar argument can be made when one examines the geometry near the D5-branes using \eqref{exactTSUN} before taking the large $N$ limit.

Due to the presence of N5-branes, the second issue for our calculation is to understand if the string coupling, $g_s$, is small so that string loop corrections can be ignored.  The dilaton in the central region, $-1 < x < 1$, is given by
\begin{align}
g_s = e^{2\phi} &= \bigg( \frac{2N}{\pi} \bigg)^{-x} \, \sqrt{\frac{2+\cos(2y)}{2-\cos(2y)}} \ .
\end{align}
We observe that the dilaton is small in the region $0 < x < 1$ but is big in the region $-1 < x < 0$.  We first focus our attention on the region $0 < x < 1$.  In the large $N$ limit, the string coupling is small except in the neighborhood of $x = 0$, where it is of order one.  Thus we expect string loop corrections to be important, but again we argue that since they are localized near $x=0$, they will give contributions at most of order $N^2$.

For the region $-1 < x < 0$, we find that the string coupling is generically large and one might expect string loop corrections to modify the leading $N^2 \ln N$ behavior.  From this point of view, the exact match between gravity and CFT partition functions is surprising and we do not have a good a priori argument for why string loop corrections do not modify the $N^2 \ln N$ behavior.  One possible explanation can be given in terms of a local S-duality transformation in this region.  To be more precise, we divide the manifold into three regions $-1 < x < -\epsilon$, $-\epsilon < x < \epsilon$ and $\epsilon < x < 1$ with $\epsilon \ll 1$.  In the first region, we make an S-duality transformation, while in the third region the theory is already weakly coupled.  The middle region then has to interpolate between two different S-duality frames and we do not know how to compute the action there.  However, since the $\ln N$ enhancement requires the entire internal space and patching only needs to occur locally in the region near $x=0$, one might hope that the middle region does not receive the $\ln N$ enhancement.  Of course this argument is only heuristic and it would be interesting to either make it more precise or determine the exact mechanism for why the loop corrections are suppressed.  Similar situations arise when one examines the geometry near the NS5-branes using \eqref{exactTSUN} before taking the large $N$ limit.

\subsection{$T^{\rho}_{\hat{\rho}}[SU(N)]$}

We now consider more general partitions which take the form \eqref{rhogeneral}.
In this case, there is a single NS5-brane stack and the charge relations, (\ref{ginvN3}), can be easily inverted to express the phases $\delta_a$ and $\hat \delta$ in terms of the partitions $\rho$ and $\hat \rho$:
\begin{align}
\delta_a - \hat \delta &= - \ln \left[\tan \left( \frac{\pi}{2}
 \frac{l^{(a)}}{\hat N_5} \right)\right]  \ .
\end{align}
To analyze the large $N$ behavior, we proceed analogously to the $T[SU(N)]$ case and consider the limit where $\hat \delta \rightarrow - \infty$ and the $\delta_a \rightarrow \infty$.  In this case, we approximate the harmonic functions by the following expressions
\begin{align}
\begin{aligned}
h_1 &= \alpha' \sin(y) \sum_{a=1}^p N_5^{(a)} e^{x-\delta_a} + ...& &\textrm{if}\ \ x < \delta_1\ ,\cr
    &= \alpha' \sin(y) \sum_{a=i}^p N_5^{(a)} e^{x-\delta_a} + ...& &\textrm{if}\ \ \delta_i < x
 < \delta_{i+1} \ , \cr
h_2 &= \alpha' \cos(y) \hat N_5 e^{-(x - \hat \delta)} +
 ...& &\textrm{if}\ \ x>\hat \delta \ ,
\end{aligned}
\end{align}
while the regions with $x > \delta_p$ and $x<\hat \delta$ will give only subleading contributions.
In this approximation we find that $W = - h_1 h_2$ so that
\begin{align}
\begin{aligned}
-W h_1 h_2  &=  \frac{1}{4} (\alpha')^4 \hat N_5^2 \left( \sum_{a=1}^p
 N_5^{(a)} e^{-(\delta_a - \hat \delta)} \right)^2 \sin^2(2y)&
&\textrm{if}\ \ \hat{\delta}<x < \delta_1 \ ,\cr
&=  \frac{1}{4} (\alpha')^4 \hat N_5^2 \left( \sum_{a=i}^p N_5^{(a)}
 e^{-(\delta_a - \hat \delta)} \right)^2 \sin^2(2y)& &\textrm{if} \ \
 \delta_i < x < \delta_{i+1} \ .
\end{aligned}
\end{align}
Using this in \eqref{vol6} we find
\begin{align}
\textrm{vol}_6&=32(4\pi)^2 \int_{\hat \delta}^{\delta_p}\! dx \int_0^{\frac{\pi}{2}}
\! dy \,(-W h_1 h_2) \nonumber \\
&=
32 \pi^3 (\alpha')^4 \hat{N}_5^2 \sum_{i=1}^p \left( \sum_{a=i}^p N_5^{(a)} e^{-(\delta_a - \hat \delta)} \right)^2 (\delta_{i} - \delta_{i-1})
\end{align}
where we define $\delta_0 \equiv \hat \delta$.  Plugging into \eqref{acteff} and combining all of the numerical factors, we obtain
\begin{align}
S_{\rm eff} &=  \frac{2}{\pi^2} \hat N_5^2 \sum_{i=1}^p \left(
 \sum_{a=i}^p N_5^{(a)} e^{-(\delta_a - \hat \delta)} \right)^2
 (\delta_{i} - \delta_{i-1}) + ... \ .
\end{align}

We now consider the scaling behavior defined by \eqref{scaling}.  The idea is to introduce separations between the $\delta_a$ which are of order $\ln N$.  In this case each region between a given $\delta_a$ and $\delta_{a+1}$ will contribute to the action at order $N^2 \ln N$.  In terms of this scaling the action becomes
\begin{align}
S_{\rm eff}
&= \frac{1}{2} N^2 \left[  \left( \sum_{a=1}^p \gamma_a \lambda^{(a)} \right)^2 \ln \left( \frac{2}{\pi} \frac{\hat \gamma}{l^{(1)}} N \right)
+ \sum_{i=2}^p \left( \sum_{a=i}^p \gamma_a \lambda^{(a)} \right)^2 \ln \left( \frac{l^{(i-1)}}{l^{(i)}} \right) \right]
+ {\cal O}(N^2) \ ,\nonumber \\
&=  \frac{1}{2} N^2\ln N \left[  (1-\kappa_1)
+ \sum_{i=2}^p \left( \sum_{a=i}^p \gamma_a \lambda^{(a)} \right)^2\left( \kappa_{i-1}-\kappa_i\right)
\right] + {\cal O}(N^2) \ ,
\end{align}
which coincides with \eqref{Fgeneral}.

\subsection{Subleading Terms}\label{sec.sublead}

So far we have concentrated on the leading $N^2 \ln N$ contributions to the free energy and it is a natural question to ask about the subleading $N^2$ contributions.  Comparing the CFT and gravity partition functions, we find that the subleading $N^2$ contributions do not match.\footnote{We have checked this numerically for $T[SU(N)]$ using the full expressions for the harmonic functions \eqref{exactTSUN}.}  However, this is not surprising since the gravity solution contains 5-brane singularities around which supergravity approximation breaks down.  Additionally, there are regions in the bulk of $\Sigma$ where the string coupling becomes large.  It would be interesting to interpret and if possible match the subleading contributions to the CFT partition function with higher curvature corrections, coming from both string and loop corrections, on the gravity side.  For the $T[SU(N)]$ theory, we note that near the D5-brane singularity, the Ricci scalar, \eqref{riccscl} does not depend on $N$ and so all powers of $R$ will contribute at order $N^2$.  Similarly, one may check that other contractions of the Riemann tensor will also contribute at order $N^2$.  Thus even at order $N^2$, the CFT partition function contains information about all orders of the higher curvature corrections.

\section*{Acknowledgments}
The authors would like to thank C.~Bachas and J.~Gomis collaboration
in early stages of this project, and also for inputs crucial for the
completion of this project.
J.~E. and M.~Y. would like to thank Simons Center for Geometry for Physics
(Simons Summer Workshop in Mathematics and Physics 2011) for hospitality.
J.~E. is supported by the FWO - Vlaanderen, Project No. G.0651.11, and by
 the ``Federal Office for Scientific, Technical and Cultural Affairs through the Inter-University Attraction Poles
 Programme,''  Belgian Science Policy P6/11-P, as well as the European Science Foundation Holograv
Network.

\appendix

\section{Barnes $G$-function}\label{app.Barnes}

Let us briefly summarize the properties of the Barnes $G$-function.
Barnes $G$-function $G_2(z)$ satisfies
\beq
G_2(z+1)=\Gamma(z) G_2(z), \quad G_2(1)=1 \ .
\eeq
From the definition it follows that
\beq
G_2(N)=(N-2)!(N-3)!\cdots 1!, \quad N=2,3,\cdots \ .
\eeq
Its asymptotic expansion is given by
\begin{equation}
\label{Gasymptotics}
\ln G_2(N+1)=\frac{N^2}{2} \ln N - \frac{3}{4} N^2+\mathcal{O}(N) \ .
\end{equation}

\section{Flux Formulas}\label{app.Fluxes}

The NS-NS and R-R three forms can be written as
\begin{align}
\label{3forms0}
\textrm{\underline{3-forms:}}\qquad H_{(3)}  =   \omega^{\, 45}\wedge db_1  \qquad {\rm and} \qquad  F_{(3)} =   \omega^{\, 67}\wedge db_2   \ ,
\end{align}
where $ \omega^{\, 45}$ and $ \omega^{\, 67}$ are the volume forms of the unit-radius  spheres  $S_1^{2}$ and $S_2^{2}$, while the gauge potentials $b_1$ and $b_2$ are given by
\begin{align}
\label{3forms1}
b_1 &= 2 \imath h_1 {h_1 h_2 (\p  h_1\bar  \p  h_2 -\bar \p  h_1 \p  h_2) \over N_1} + 2  h_2^D \ ,  \cr
b_2 &= 2 \imath h_2 {h_1 h_2 (\p  h_1 \bar\p  h_2 - \bar\p  h_1 \p  h_2) \over N_2} - 2  h_1^D \ .
\end{align}
In this expression one needs the dual harmonic functions, defined by
\begin{align}
h_1 = -\imath ({\cal A}_1 - \bar {\cal A}_1) \qquad& \rightarrow \qquad h_1^D = {\cal A}_1 + \bar {\cal A}_1 \ , \cr
h_2 = {\cal A}_2 + \bar {\cal A}_2\qquad  \qquad& \rightarrow \qquad h_2^D = \imath ({\cal A}_2 - \bar {\cal A}_2) \ .
\end{align}
for holomorphic harmonic functions $\mathcal{A}_1, \mathcal{A}_2$.  The constant ambiguity in the definition of the dual functions is related to changes of the background fields under large gauge transformations.
The expression for the gauge-invariant self-dual 5-form is given by:
\begin{align}
\label{5form0}
\textrm{\underline{5-form:}}\qquad F_{(5)}  = - 4\,  f_4^{\, 4}\,  \omega^{\, 0123} \wedge {\cal F} + 4\, f_1^{\, 2}f_2^{\, 2} \,  \omega^{\, 45}\wedge \omega^{\, 67}
 \wedge (*_2  {\cal F})   \ ,
\end{align}
where $ \omega^{\, 0123}$ is the volume form of the unit-radius ${\rm AdS}_4$,
${\cal F}$ is a 1-form on $\hat{\rho}$ with the property that $f_4^{\, 4} {\cal F}$ is closed,
and $*_2 $ denotes Poincar\' e duality with respect to the $\hat{\rho}$ metric.
The explicit expression for ${\cal F}$  is given by
\begin{align}\label{calF}
f_4^{\, 4} {\cal F} = d j_1\   \qquad {\rm with} \qquad j_1 =
3 {\cal C} + 3 \bar  {\cal C}  - 3 {\cal D}+ \imath \frac{h_1 h_2}{W}\,   (\p  h_1 \bar\p  h_2 -\bar \p h_1 \p h_2) \ ,
\end{align}
where ${\cal C}$ and ${\cal D}$ are defined by  $\p {\cal C} = {\cal A}_1 \p  {\cal A}_2 - {\cal A}_2 \p  {\cal A}_1$
and  ${\cal D} = \bar {\cal A}_1 {\cal A}_2 + {\cal A}_1 \bar {\cal A}_2$.


\bibliographystyle{utphys}
\small\baselineskip=.8\baselineskip

\end{document}